\documentclass[twocolumn,prb,printnumbers,superscriptaddress,amsmath,amssymb]{revtex4}
\usepackage{graphicx}
\usepackage{color}

\begin{document}
\title{Sound attenuation  in  finite-temperature stable glasses}
\date{\today}

\author{Lijin Wang}
\affiliation{School of Physics and Materials Science, Anhui University, Hefei 230601, P. R. China.}
\author{Elijah Flenner}
\affiliation{Department of Chemistry, Colorado State University, Fort Collins, Colorado 80523, USA}
\author{Grzegorz Szamel}
\affiliation{Department of Chemistry, Colorado State University, Fort Collins, Colorado 80523, USA}

\begin{abstract}
The temperature dependence of the thermal conductivity of amorphous solids  is markedly different from that of their crystalline counterparts, but exhibits universal behaviour.
Sound attenuation is believed to be related to this universal behaviour.
Recent computer simulations demonstrated that in the harmonic approximation
sound attenuation $\Gamma$ obeys quartic, Rayleigh scattering scaling for small wavevectors $k$ and
quadratic scaling for wavevectors above the Ioffe-Regel limit. However,
simulations and experiments do not provide a clear picture of what to expect at finite temperatures where anharmonic
effects become relevant.  Here we study sound attenuation at finite temperatures for model glasses of various stability, from
unstable glasses that exhibit rapid aging to glasses whose stability is equal to those created in laboratory experiments.
We find several scaling laws
depending on the temperature and stability of the glass. First, we find the large wavevector quadratic scaling to
be unchanged at all temperatures. Second, we find that at small wavectors $\Gamma \sim k^{1.5}$ for an aging glass,
but $\Gamma \sim k^2$ when the glass does not age on the timescale of the calculation.
For our most stable glass, we find that $\Gamma \sim k^2$ at small wavevectors, then a crossover to Rayleigh scattering scaling $\Gamma \sim k^4$, followed by
another crossover to the quadratic scaling at large wavevectors.
Our computational observation of this quadratic behavior reconciles simulation, theory and experiment,
and will advance the understanding of the temperature dependence of thermal conductivity of glasses.
\end{abstract}

\maketitle

\section{Introduction}

Glasses exhibit  universal low-temperature thermal properties that differ markedly from those of their crystalline counterparts~\cite{Pohl2002,Zeller1971, Zaitlin1975}.
For instance, the thermal conductivity of glasses for temperatures $T$ below approximately 1K increases as $T^2$ instead of $T^3$ as for crystalline solids \cite{Pohl2002}.
Additionally, there is a plateau in the thermal conductivity of glasses around 10K that is absent in crystalline solids.
While the two-level tunneling model~\cite{Zeller1971,Phillips1972,Anderson1972,Pohl2002,Lubchenko2003,Lubchenko2018,Lubchenko2001} predicts the $T^2$ increase of the thermal conductivity of glasses, there needs to be another effect
to account for the $\sim 10$K plateau.
Zeller \textit{et al.}\ \cite{Zeller1971} argued that two-level tunneling combined with sound waves where sound attenuation $\Gamma$
obeys Rayleigh scattering scaling,
$\Gamma \sim k^4$, at small wavevector $k$ correctly predict the low temperature $T^2$ and the plateau of the thermal conductivity.
However, their calculation did not provide direct evidence of small wavevector Rayleigh scattering scaling of sound attenuation.

Scattering experiments were able to provide direct evidence for the existence of Rayleigh scattering scaling
at low temperatures \cite{Rayleigh_RufflePRL2006,Rayleigh_MonacoPNAS2009_exp,Rayleigh_BaldiPRL2010,Rayleigh_BaldiPRL2014,Rayleigh_BaldiPRL2013,Rayleigh_RutaJCP2012,MasciovecchioPRL2006},
and how anharmonicity modifies sound attenuation at finite temperatures \cite{Rayleigh_BaldiPRL2014}.
Baldi \textit{et al.}\ \cite{Rayleigh_BaldiPRL2014} used inelastic x-ray scattering to demonstrate that the small wavector, low temperature
sound attenuation obeys Rayleigh scattering scaling for a network glass former. However, they also found that at higher temperatures sound
attenuation scales quadratically with wavevector at small wavevectors and at large wavevectors with a quartic,
Rayleigh scattering scaling region in between. The small wavevector quadratic scaling was strongly temperature dependent, while
the large wavevector quadratic scaling was nearly temperature independent.  Other scattering studies support the picture
that anharmonic effects give rise to $\Gamma \sim k^2$ for both small and large wavevecctors for most glass formers
\cite{MasciovecchioPRL2004,BenassiPRB2005,ScopignoPRL2006,Devos2008}.  The small wavevector
quadratic scaling has been attributed to damping due to a spatial variation of the strain field leading to heat flow (thermoelastic dissipation)
\cite{Zener1938,Lifshitz2000,De2006},
heat flow between different sound modes (Akhiezer damping) \cite{Akhiezer1,Akhiezer2}, and to spatial fluctuations of the
elastic constants (fluctuating elasticity) \cite{Schirmacher2010PRB}.

Surprisingly, it was
found in vitreous germanium that sound attenuation increased linearly with frequency at small frequencies \cite{FerranteMarginalStability2013}, implying a
linear instead of quadratic increase of sound attenuation with wavevector and non-universal
behavior. Fluctuating elasticity theory
predicts that anharmonic effects results in non-quadratic scaling of sound attenuation at small wavevectors close to an elastic instability \cite{MarruzzoMarginalStability2013},
and a quadratic scaling of sound attenuation sufficiently far from an elastic instability \cite{Tomaras2010}. Additionally, fluctuating elasticity theory predicts that
the harmonic quartic scaling at small wavevectors is observable at low enough temperatures  \cite{Tomaras2010}. While fluctuating elasticity theory
correctly describes sound attenuation qualitatively, it has recently been shown to fail quantitatively \cite{Caroli2019}.

Recently, large-scale computer simulations confirmed Rayleigh scattering scaling of sound attenuation in the
harmonic approximation at small wavevectors and quadratic scaling at large wavevectors \cite{Mizuno2018,Wang2019SMattenuation,Lerner2019JCP}.
It is not
expected that the quadratic small wavevector scaling could be captured in the harmonic approximation since this
scaling is attributed to anharmonic effects. Between the small wavevector quartic and large wavevector quadratic regime a
possible crossover $k^4 \ln(k)$ regime has been identified in simulations \cite{LemaitreNatMat2016,Caroli2019,Lerner2019JCP,Mizuno2018,Wang2019SMattenuation},
but this crossover region shrinks with increasing glass stability~\cite{Wang2019SMattenuation}.
To observe Rayleigh scattering scaling in simulations researchers have to utilize very large systems\cite{Mizuno2018} or examine very stable glasses\cite{Wang2019SMattenuation}.
These stable simulated glasses have only  recently become available due to the combination of the swap algorithm and model polydisperse
glass formers \cite{Berthier2016PRL,Ninarello2017PRX}. The Rayleigh scattering scaling regime extends to larger wavevectors as the stability of the
glass increases, and thus smaller systems can be simulated to clearly observe Rayleigh scattering scaling \cite{Wang2019SMattenuation}.

Simulations have not provided a clear picture of the temperature dependence of sound attenuation in model glass formers. Busselez, Pezeril, and Gusev \cite{BusselezJCP}
used molecular dynamics simulations to examine the sound attenuation in a model of glycerol.  For small frequencies, they reported quadratic scaling
of sound attenuation with
frequency at intermediate temperatures and cubic scaling with frequency at lower temperatures. They did not clearly see the three regimes
observed in the experiments of Baldi \textit{et al.} \cite{Rayleigh_BaldiPRL2014}, the
low and high frequency quadratic scaling and the intermediate quartic scaling.
Mizuno and Mossa recently studied glasses obtained by rapidly quenching a mono-disperse Lennard-Jones fluid \cite{Mizuno2019CondesMatterFiniteT,Mizuno2019arxivFiniteT},
and found results consistent with fluctuating elasticity theory
close to an elastic instability \cite{MarruzzoMarginalStability2013}.  Specifically, at a finite temperature below the glass transition temperature but high enough for
anharmonic effects to be present, Mizuno and Mossa found that $\Gamma \sim k^{1.5}$ at small wavevectors, $\Gamma \sim k^2$ at large wavevectors,
and an intermediate quartic regime, $\Gamma \sim k^4$.

The collection of these results leaves an unclear picture of the small wavevector sound attenuation in finite temperature amorphous solids.
Here we are able to help clarify the picture by studying the temperature dependence of sound
attenuation in poorly annealed glasses and in extremely stable glasses.
The stability of our stable glasses is comparable to that of exceptionally stable laboratory glasses created using
vapor deposition~\cite{SwallenScience2007,BerthierPRL2017vapor}.
We observe nearly all the scaling behavior reported in the experiments and simulations discussed above.
For our poorly annealed glasses, we find that $\Gamma \sim k^{1.5}$ at small wavevectors
while the glass is undergoing aging on the time scale of the simulation. The evidence for aging comes from an upturn in the mean-square-displacement.
For glasses that are not noticeably aging on the time scale of the simulation, we find that $\Gamma \sim k^2$ at small wavevectors and large
wavevectors. Between these two quadratic scaling regimes we clearly observe a quartic scaling regime in our extremely stable glasses. We characterize the
temperature dependence of the different scaling regimes.

\section{Simulation details}
We simulated $N=\{48000, 96000,192000\}$ polydisperse spheres having equal mass $m$ with periodic boundary conditions.
Particle diameters $\sigma\in [0.73,1.63]$ have a distribution $P(\sigma)\sim  \sigma^{-3}$.
To prevent demixing, we employ a non-additive mixing rule to determine the cross-diameter $\sigma_{ij}$,
$\sigma_{ij}=\frac{\sigma_{i}+\sigma_{j}}{2}(1-\epsilon |\sigma_{i}-\sigma_{j}|)$ with $\epsilon=0.2$.
Particles ${i}$ and ${j}$  interact via
the inverse power law potential $ V(r_{ij}) = \left(\frac{\sigma_{ij}}{r_{ij}}\right)^{12} + V_{c}(r_{ij})$
when the separation between particles $i$ and $j$, $r_{ij} < r_{ij}^c=1.25\sigma_{ij}$ and
$ V(r_{ij})=0$ if $r_{ij} \geq r_{ij}^c$. Here,  $V_{c}(r_{ij})=c_{0}+c_{2}\left(\frac{r_{ij}}{\sigma_{ij}}\right)^{2}
+ c_{4}\left(\frac{r_{ij}}{\sigma_{ij}}\right)^{4}$ and the coefficients are chosen so that $V(r_{ij})$ and its first two derivatives are continuous at
$r_{ij}^c$.
The number density $\rho=1.0$. For reference,
the  onset temperature of slow dynamics  $T_o\approx 0.200$, the mode coupling temperature $T_c \approx 0.108$, and
the estimated experimental glass temperature $T_g \approx 0.072$~\cite{Ninarello2017PRX,Wang2019NC}.

We obtain  equilibrated supercooled liquids using the swap Monte Carlo algorithm~\cite{Berthier2016PRL,Ninarello2017PRX}
at temperatures ranging from $T_o$ down to  $0.062$, which is lower than $T_g$.
To create a glass, we quench a configuration equilibrated at a temperature $T_p$, which we call the parent temperature of this glass,
to its inherent structure  using the fast inertial relaxation engine minimization~\cite{fire}.
We then heat the glass to the desired temperature $T$ in the $NVT$ ensemble~\cite{lammps1,lammps2}. Finally,  the system is equilibrated
for a time referred in the following as the
ageing time $\tau_{age}$  before the production runs start.
The total length for the production run is equal to $4000$, and we average over different initial configurations equilibrated at $T_p$.
Thus, the finite temperature glass is characterized by three parameters \{$T$,$T_p$,$\tau_{age}$\}
and its stability is determined by a combination of $T_p$ and $\tau_{age}$.

We calculate sound attenuation $\Gamma_{\lambda}$ from the decay of the current density correlation functions
\begin{equation}
 C_{\lambda} (k, t) =\Bigg\langle
\frac{ \vec{J}_{\lambda}(k, t) \cdot \vec{J}_{\lambda} (-k, 0) }
{ \vec{J}_{\lambda}(k, 0) \cdot \vec{J}_{\lambda} (-k, 0) }\Bigg\rangle
\label{jL}
\end{equation}
with
\begin{equation}
\vec{J}_T (k, t) =
\sum_{j=1}^{N} [\vec{v}_j(t) - (\vec{v}_j(t)\cdot \hat{k}) \hat{k}]
e^{i\vec{k}\cdot \vec{r}_j(t)}\label{jT}
\end{equation}
for transverse current, $T$, and
\begin{equation}
\vec{J}_L (k, t) = \sum_{j=1}^{N} [(\vec{v}_j(t)\cdot \hat{k}) \hat{k}]
e^{i\vec{k}\cdot \vec{r}_j(t)}\label{jl}
\end{equation}
for longitudinal current, $L$.
Here, $\vec{v}_j(t)$ is the velocity of particle $j$ at time $t$, $k= |\vec{k}|$, $\hat{k}=\vec{k}/|\vec{k}|$ with $\vec{k}$ the wavevector.

\begin{figure}
\centering
\includegraphics[width=0.5\textwidth]{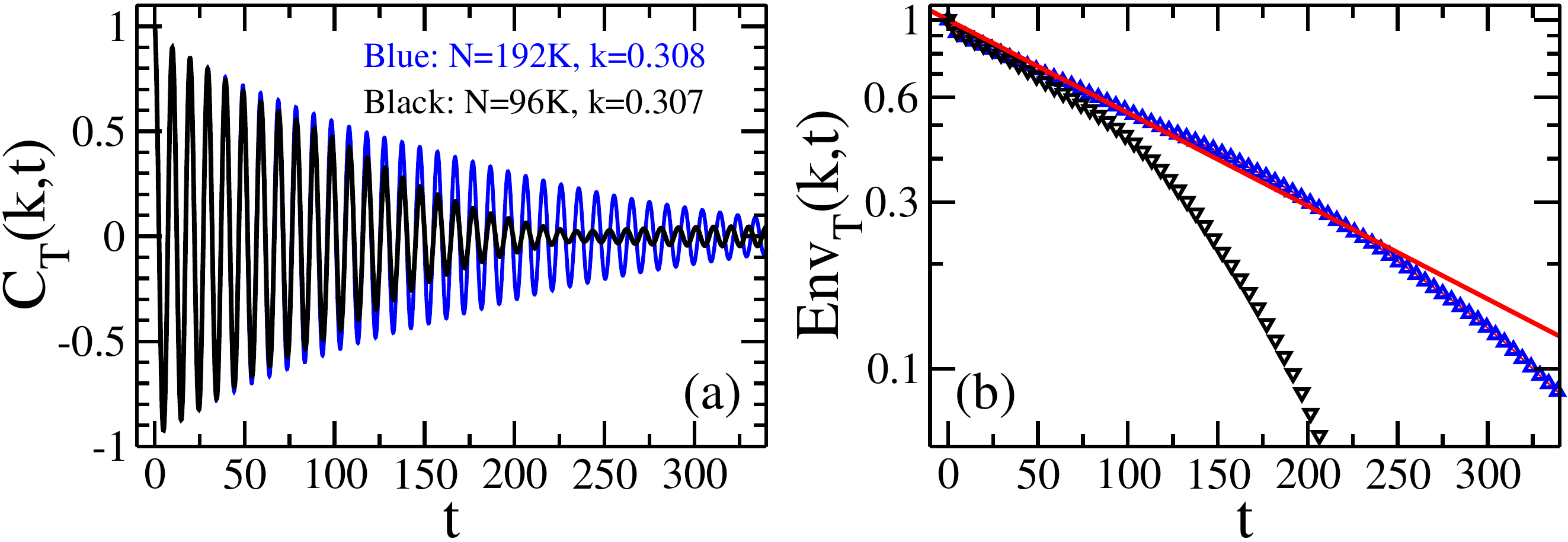}
\caption{ Transverse current correlation functions $C_T(k,t)$ in the left panel, (a), and their corresponding  envelopes $\mathrm{Env}_{T}(k,t)$
in the right panel, (b),
for two system sizes, $N=96K$ (black) and $192K$ (blue), at a similar magnitude of the wavevector $k\approx 0.31$.
The solid line in (b) represents a fit  to $\exp(-\Gamma t/2)$. The fit works well up to longer times in $N=192K$
system.}
\label{fig1}
\end{figure}

Previous studies demonstrated either explicitly~\cite{Wang2019SMattenuation,Lerner2019JCP}or implicitly~\cite{Bouchbinder2018NJP}  that there are finite size effects in
the calculations of sound attenuation within the harmonic approximation. We find the finite-size effects persist for finite temperatures, especially at low temperatures.
Shown in  Fig.~\ref{fig1}(a), are $C_{T}(k, t)$ at nearly the same wavevector but for two different system sizes.
They overlap and decay exponentially at short times, but then deviate at longer times.

 To eliminate finite size effects, we use a restricted envelope fit method as described in detail by  Wang \textit{et al.}~\cite{Wang2019SMattenuation}. We
 determine $\Gamma_{\lambda}$ by fitting the envelope of $C_{\lambda}(k, t)$ to ${\rm exp}{(-\Gamma_{\lambda} t/2)}$
 up to the time when  the envelope starts to significantly deviate from exponential decay.
 Shown in Fig.~\ref{fig1}(b) is the envelope of $C_T(k,t)$ shown in
 Fig.~\ref{fig1}(a) and a fit to the larger system.
 Fits to the exponentially decaying part give the same sound attenuation within error. Therefore, in this work sound attenuation
 combined from different system sizes is shown without distinguishing  system sizes since finite size effects have been removed.

\section{Temperature dependence of sound attentuation in stable glasses}
\begin{figure}[!t]
\centering
\includegraphics[bb=0in 3.0in 8in 5in,clip=false,width=0.46\textwidth]{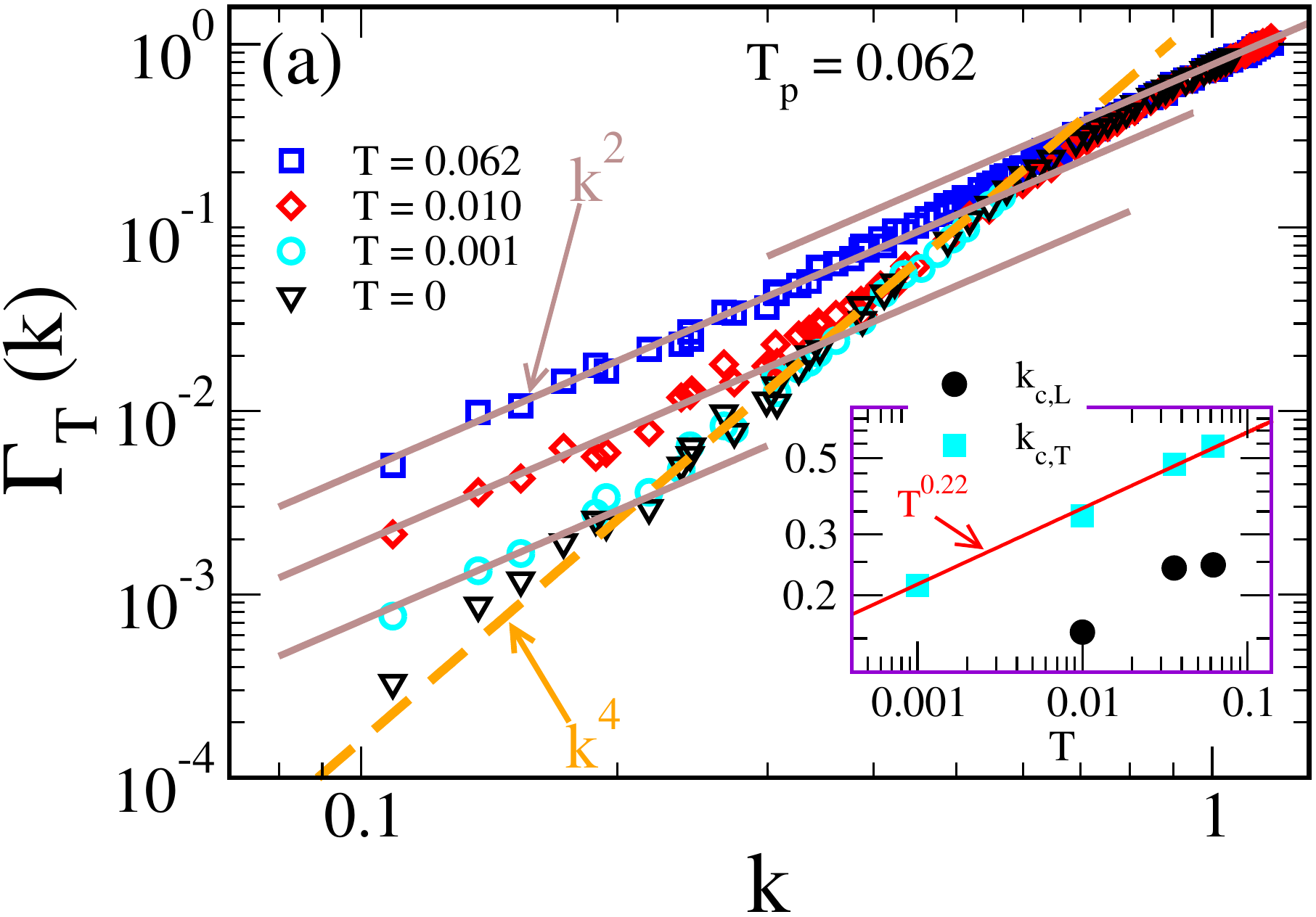}
\includegraphics[bb=0.1in -0.1in 8in 8.2in,clip=false,width=0.46\textwidth]{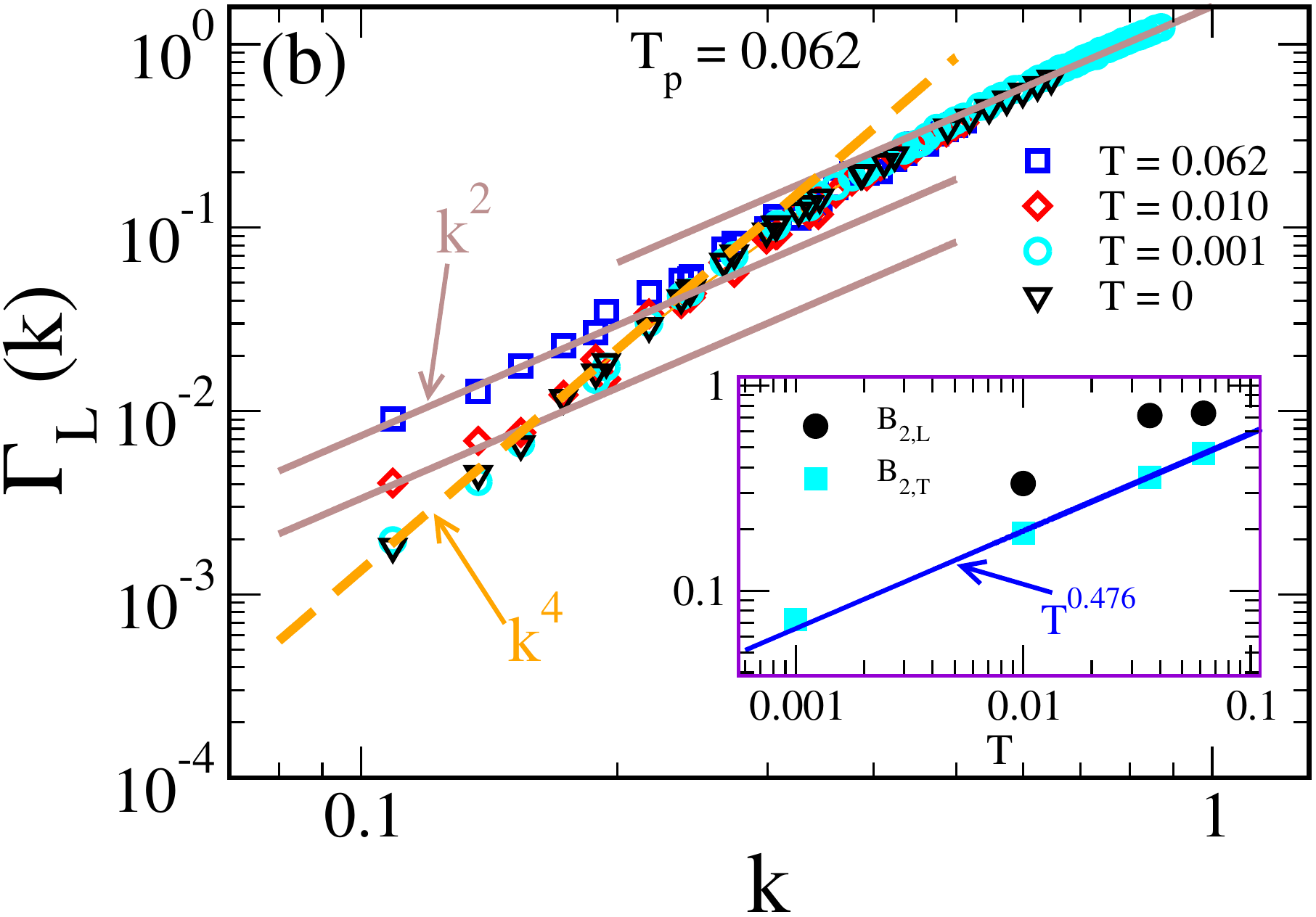}

\caption{  Wavevector $k$ dependence of transverse sound attenuation $\Gamma_T(k)$ (a) and longitudinal sound attenuation $\Gamma_L(k)$ (b)
for our most stable glass, $T_p = 0.062$ at different temperatures $T$. The aging time $\tau_{\mathrm{age}}=1000$, but these
results are independent of the aging time.
The inset in (a) shows the $T$ dependent upper wavevector $k_{c,T}$ and $k_{c,L}$  for the low-$k$ quadratic scaling.
$k_{c,\lambda}$ at each $T$ is roughly defined as the crossover between the low-$k$ quadratic law  (solid line) and the intemediate-$k$ quartic law  (dashed line)
in  (a) and (b). The solid line in the inset is a fit of $k_{c,T}$ to ${T}^{0.22}$. The inset in (b) shows the $T$ dependence of the low-$k$ quadratic coefficients
$B_{2,{T}}$ and $B_{2,L}$, \emph{i.e.} $\Gamma_T(k)=B_{2,T}k^2$ and $\Gamma_L(k)=B_{2,L}k^2$. The solid line in the inset indicates a fit of $B_{2,T}$ to ${T}^{0.476}$. }
\label{fig4}
\end{figure}

We examined the  finite-temperature sound attenuation in our most stable glasses with parent temperature $T_p=0.062$.
The stability of these model stable glasses has been demonstrated by Ninarello {\it et al.}~\cite{Ninarello2017PRX} to be
comparable to that of experimental glasses, which makes the present study of sound attenuations relevant for experimental glasses.
We find that sound attenuation in stable glasses at temperatures below $T_g \approx 0.072$
shows no $\tau_{\mathrm{age}}$ dependence in any of our simulations. We aged the systems up to $\tau_{\mathrm{age}}$ equal to one million.

Shown in Fig. \ref{fig4} is the temperature dependence of the transverse (a) and longitudinal (b) sound attenuation in our most stable glasses.
The small wavevector quadratic behaviour for $\Gamma_{\lambda}$ can be observed at all  temperatures,
except for $\Gamma_{L}$ at $T=0.001$, Fig. \ref{fig4}(b).
At small $k$, we expect that $\Gamma_L(k)\sim k^2$  at $T=0.001$ if we can simulate a much larger system.
At large wavevectors, we find a temperature independent, quadratic dependence of $\Gamma_{\lambda}$ on wavevector that is different
from the small wavevector quadratic dependence.
Rayleigh scattering scaling is found between the large wavevector and
small wavevector quadratic regimes.
The coefficients quantifying sound attenuation in the Rayleigh scattering regime, $\Gamma_{\lambda}/k^4$, are
approximately temperature-independent. We conclude that the
finite temperature anharmonic effects alter sound attenuation predominately at small wavevectors.

These results mirror the results reported in the experiments of Baldi \textit{et al.}\ \cite{Rayleigh_BaldiPRL2014}, who also reported Rayleigh
scattering scaling for
low temperatures and quadratic scaling at small wavevector with increasing temperature.  To analyze their results,
Baldi \textit{et al.}\ fit the sound attenuation to a model that included an Akhieser term and a Rayleigh scattering scaling
term.  Future work needs to examine if a similar model is reasonable for model glass formers,
but this is outside the scope of this study.

Next, we examine the temperature dependence of small-wavevector quadratic scaling of sound attenuation.
First, we study the temperature dependence of the largest wavevector, $k_{c,\lambda}$ where
$\Gamma_{\lambda}(k)$ scales quadratically,  see the inset to Fig. \ref{fig4}(a).
We define $k_{c,\lambda}$ as the intersection of the small-$k$
quadratic and intermediate-$k$ quartic fitting lines. At all temperatures, $k_{c,T}$  is found to be larger than $k_{c,L}$,
and thus, if a characteristic length scale determines this crossover,
this length scale must be different for longitudinal and transverse sound.   Moreover, as $T$ increases,  both
$k_{c,T}$  and $k_{c,L}$ increase. We find that a power law $k_{c,T}\sim T^{0.22}$ describes well the temperature dependence of $k_{c,T}$.
Given the uncertainties in our determination of
$k_{c,T}$ and the limited range of $k_{c,T}$,   we do not exclude the possibility that the temperature dependence of $k_{c,T}$
can be described by other functions.
Our fit suggests that we would need an approximately 1 million particle system to observe the small wavevector quadratic scaling
for longitudinal attenuation at $T=0.001$
for the parent temperature $T_p = 0.062$.

The temperature dependence of the small wavevector quadratic coefficients, $B_{2,\lambda}=\Gamma_{\lambda}(k)/k^2$ is shown in the inset to Fig. \ref{fig4}(b).
Both $B_{2,T}$ and $B_{2,L}$ increase with increasing $T$, and thus the small wavevector sound attenuation becomes
progressively stronger with increasing temperature. We also find  $B_{2,T}(T)$ is always smaller than $B_{2,L}(T)$,
hence the longitudinal sound attenuation is larger than the transverse wave attenuation at a fixed wavevector.
This observation is consistent with the conclusion in the study of sound attenuation within the harmonic approximation~\cite{Wang2019SMattenuation}.
Moreover, $B_{2,T}(T)$ can be fitted well with a power law, $B_{2,T}(T)\sim T^{\beta}$ with $\beta = 0.476\pm 0.015$.
This value of $\beta$ is different than the
fluctuating elasticity theory~\cite{ Schirmacher2010PRB} prediction of $\beta = 1$
for a system not close to an elastic instability.

Fluctuating elasticity theory~\cite{MarruzzoMarginalStability2013} predicts that $\Gamma_T = B_{1.5,T} k^{1.5}$ with $B_{1.5,T} \sim T^{0.5}$ close to an elastic
instability. While the temperature scaling exponent is close to the value we get from our fits, the small wavevector scaling is
different, and thus the theory is not consistent with our results. Mizuno and Mossa~\cite{Mizuno2019CondesMatterFiniteT,Mizuno2019arxivFiniteT} obtained results consistent
with fluctuating elasticity theory for a mono-disperse Lennard-Jones glass, which is a poor glass former that is prone to crystallization.
We do not observe
the $k^{1.5}$ scaling for small wavevectors for our stable glasses, but we observe this scaling for an aging glass,
which is described in Section \ref{sec:aging}.
More  work is needed  to  disentangle the  similarity and difference observed here between our simulations and existing theoretical predictions.

\begin{figure}
\centering
\includegraphics[width=0.4\textwidth]{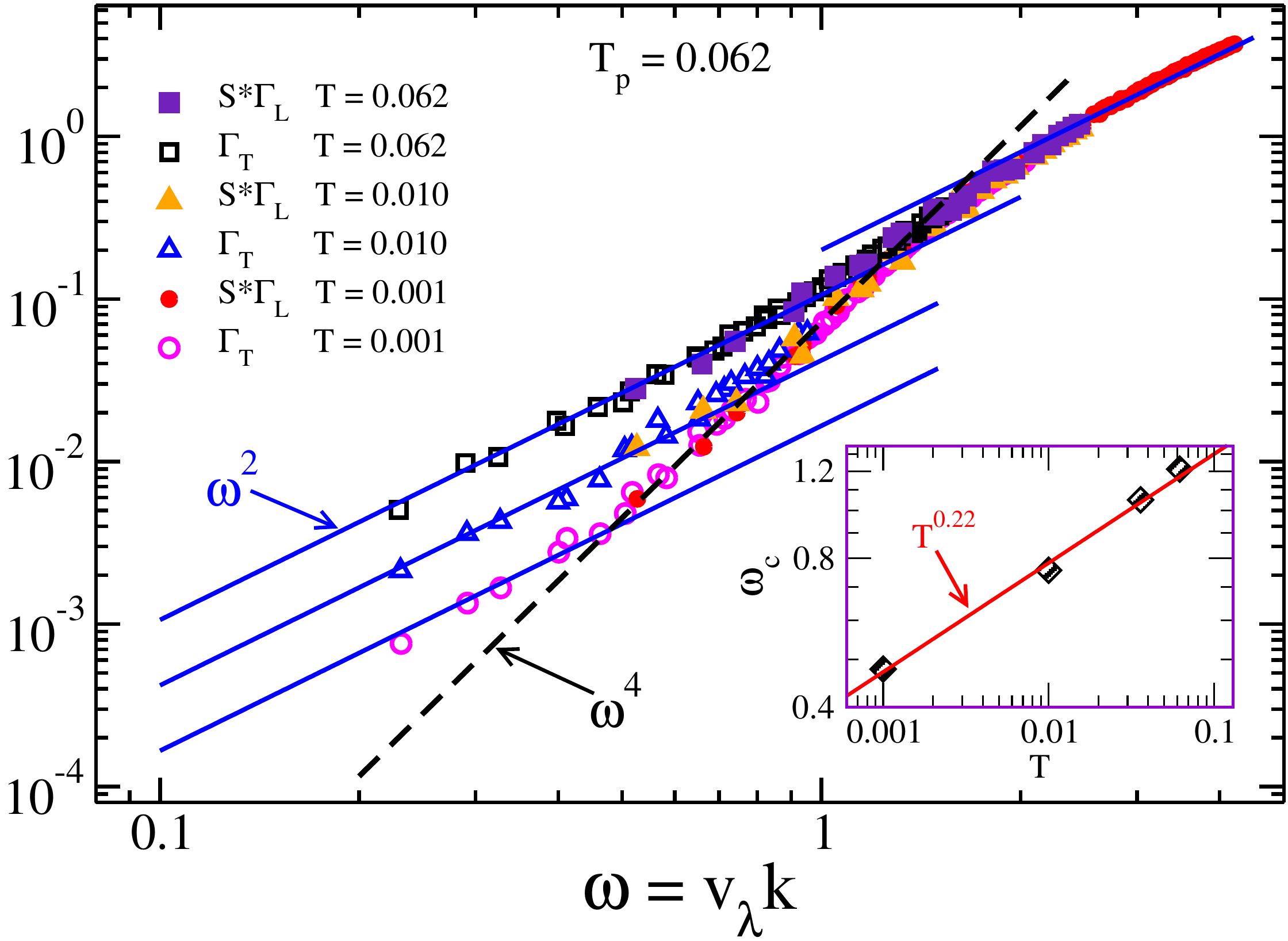}
\caption{  Sound attenuation as a function of frequency $\omega=\rm v_{\lambda}(T)k$ for glasses at $T=0.001$, $0.010$ and  $0.062$,
with parent temperature $T_p=0.062$ and  waiting time $\tau_{\mathrm{age}}=1000$.
The values  of $\rm v_{T}$ ($\rm v_{L}$) for $T=0.001$, $0.010$  and $0.062$ glasses
are approximately $2.125$ ($4.838$), $2.121$ ($4.828$) and $2.109$ ($4.811$), respectively.
The longitudinal sound attenuation  $\Gamma_L$ for each glass is multiplied by the same factor $S\approx 3$.
In the inset is the $T$ dependence of the upper frequency $\omega_{c}$  for the low-$\omega$ quadratic scaling.
We roughly define $\omega_{c}$  as the intersection of the low-$\omega$ quadratic  and the intermediate-$\omega$ quartic lines indicated in the main plot. }
\label{fig5}
\end{figure}

Previous simulation studies ~\cite{Rayleigh_MonacoPNAS2009_sim,Wang2019SMattenuation} of sound attenuation  within the harmonic approximation
demonstrated that for a fixed stability, when atenuation is examined as a function of frequency, the scaling behavior for longitudinal and transverse attenuation is the same, \textit{i.e.}\ the frequency dependent longitudinal and the transverse attenuation overlap when scaled by a constant factor.
We find that this also holds at finite temperatures, and $\Gamma_T(\omega) = S*\Gamma_L(\omega)$ where
$S \approx 3$ if when we use speeds of sound to replace the wavevector by the frequency,
$\omega = v_{\lambda} k$ where $v_\lambda$ is the speed of sound, Fig.~\ref{fig5}.
If we obtain $\omega$ directly from the fits to $C_\lambda(k,t)$ we find $S \approx 3.5$, but the scaling still holds.
At small $\omega$, we find $\Gamma_T(\omega)\sim \omega^2$ for all temperatures including $T=0.001$.
While for $T=0.001$ we do not observe $\Gamma_L(k) \sim k^2$ at small $k$, Fig.~\ref{fig4}(b),
we hypothesis that $\Gamma_L(k) \sim k^2$ would  be observed  at
smaller wavevectors than we have available due to the frequency scaling.
Additionally, since $\Gamma_T(\omega)\sim \Gamma_L(\omega)$ then the upper frequency $\omega_c$
 for this quadratic scaling of $\Gamma_{\lambda}(\omega)$ is independent of the polarization. The same
 characteristic frequency can be associated with the change of scaling of both longitudinal and transverse
 sound attenuation.
 The temperature dependence of $\omega_c$ is given in the inset to Fig. \ref{fig5}, and we find that $\omega_c$ increases as $T$ increases and scales approximately
 as $T^{0.22}$ which is the same temperature scaling as that of $k_{c,T}$.

\section{Stability Dependence of Sound Attenuation}
\label{sec:aging}
\begin{figure}[!t]
\centering
\includegraphics[width=0.38\textwidth]{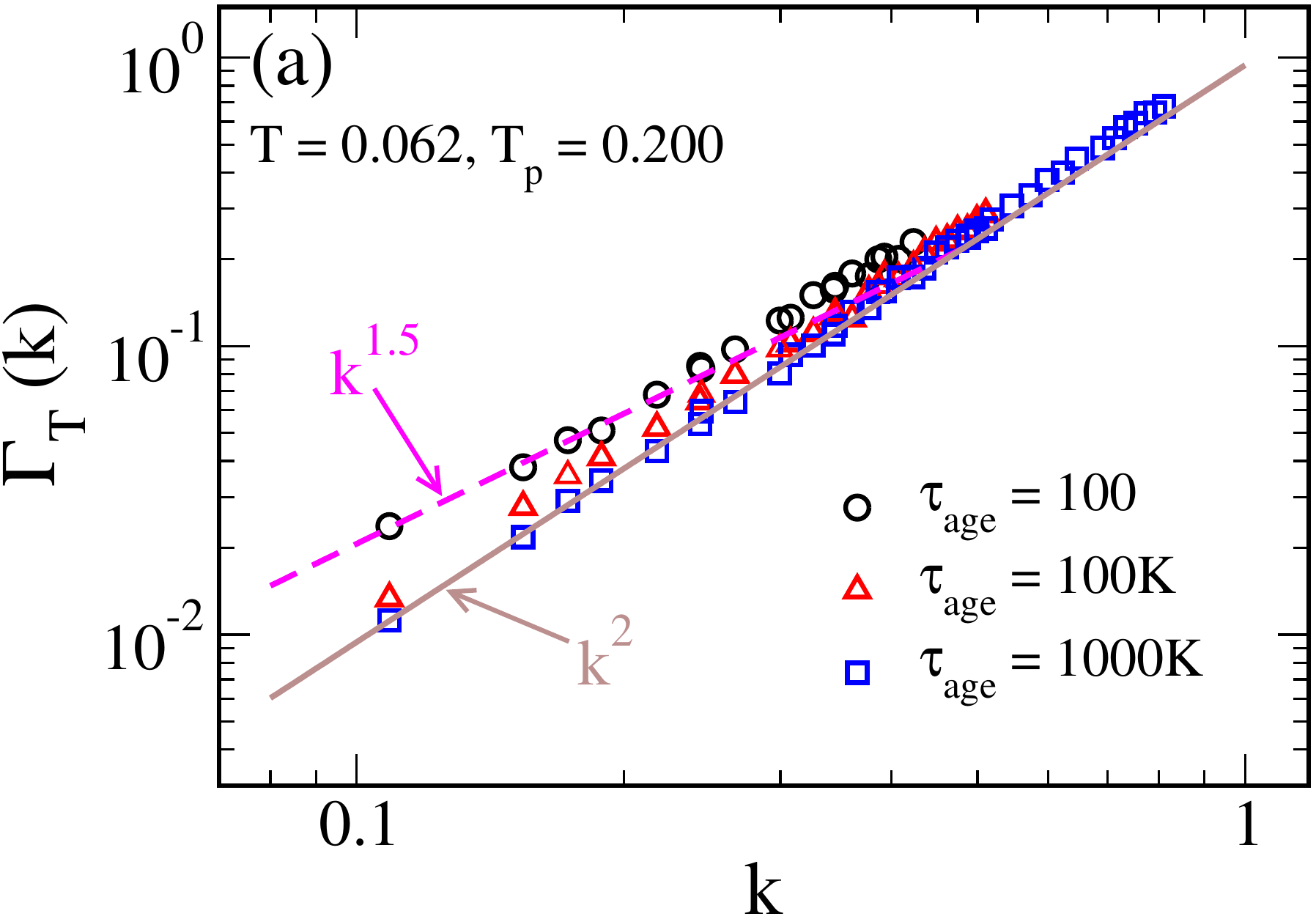}
\includegraphics[width=0.38\textwidth]{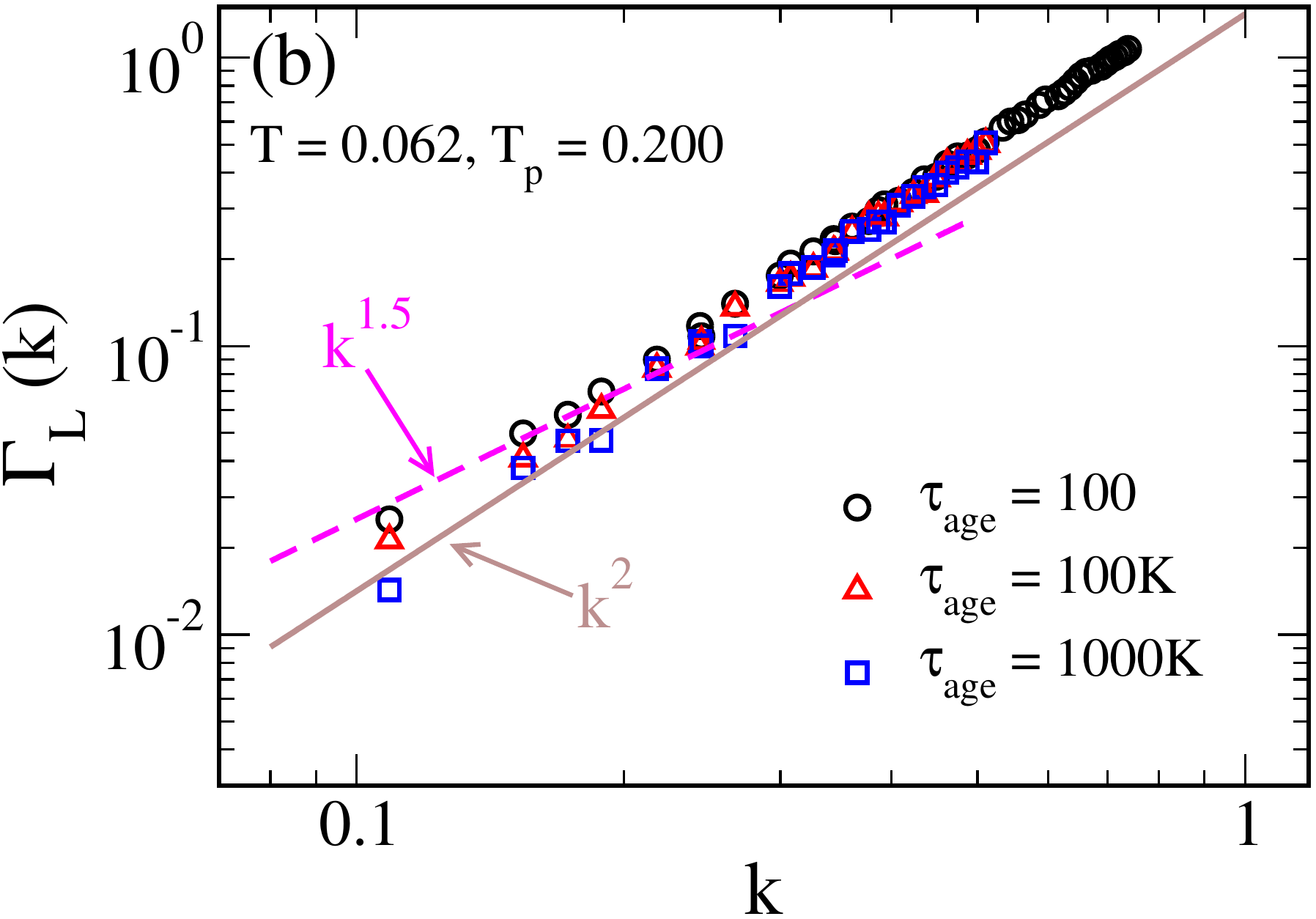}
\includegraphics[width=0.38\textwidth]{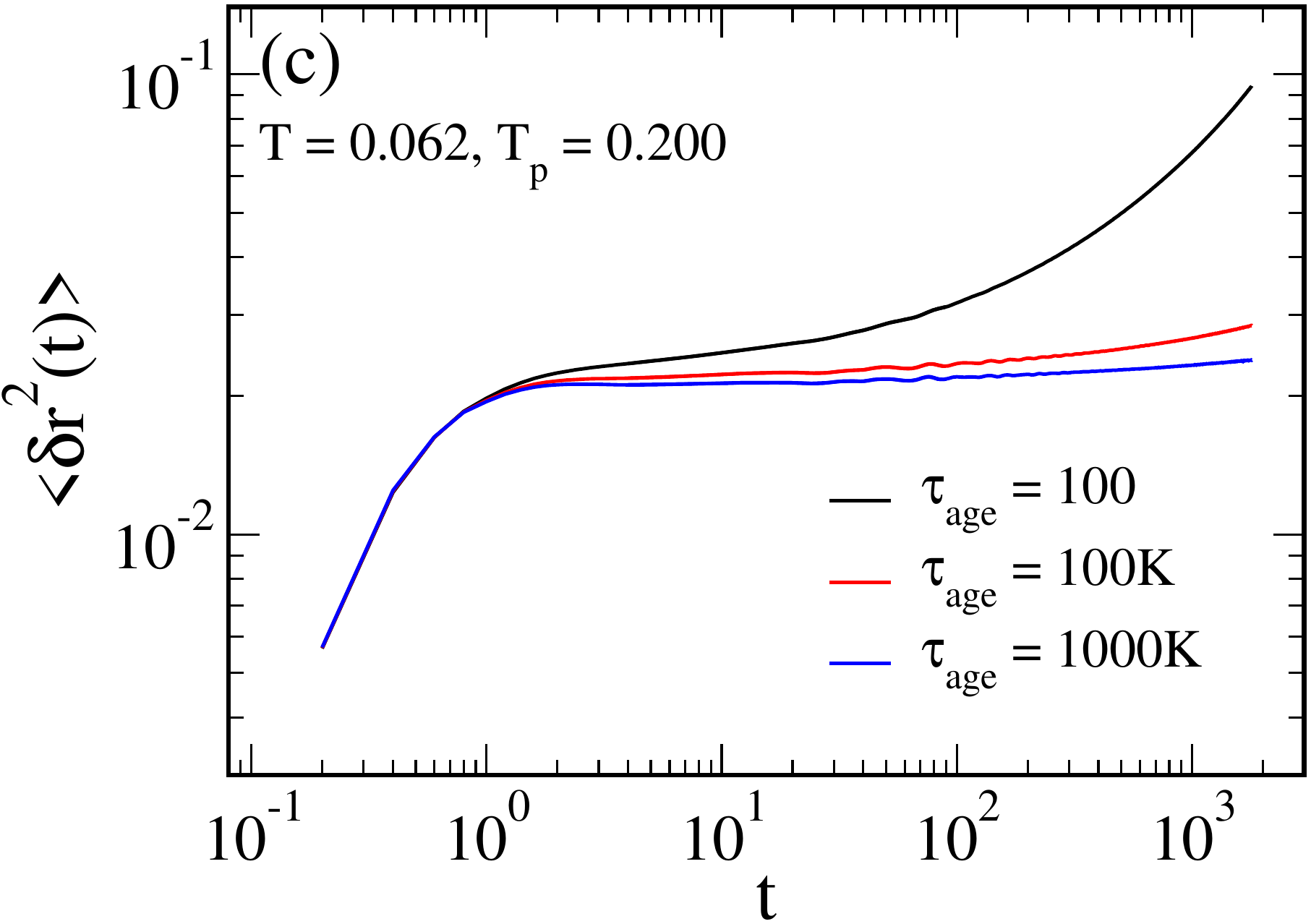}
\caption{
Transverse sound attenuation $\Gamma_T(k)$ (a) and longitudinal sound attenuation $\Gamma_L(k)$ (b) in glasses at temperature $T=0.062$
prepared by quenching from parent temperature $T_p=0.200$,
for aging times $\tau_{\mathrm{age}}=\{100,100K,1000K\}$.
The scaling of  sound attenuation changes from  $k^{1.5}$ when $\tau_{\mathrm{age}}=100$
to $k^2$ when $\tau_{\mathrm{age}}=100K$ and $1000K$. The mean-squared displacement $\left< \delta r^2(t)\right>$ (c)
dependens significantly on the aging time.  }
\label{fig2}
\end{figure}

We examined the stability dependence of sound attenuation. The stability of the glass can be
increased by aging the glass or by changing the protocol used to prepare it. For example,
the supercooled liquid can be cooled slower to create a more stable glass, or, for some glass formers, physical vapor deposition
can make very stable glasses. We begin our study of the stability dependence
of sound attenuation by examining sound attenuation in an aging glass, where we monitor the aging process
using the mean square displacement. We only study aging effects for a poorly annealed glass.
Next, we study increasingly more stable glasses by investigating the parent temperature
dependence of sound attenuation.

Our poorly annealed glass, with parent temperature
$T_p = 0.2$, is prepared by quenching from the temperature corresponding to the onset of slow dynamics.
The aging of this glass is clearly seen in its the mean square displacement
$\left< \delta r^2(t)\right>=N^{-1}\left<\sum_{j=1}^{N}[\vec{r}_j(t) - \vec{r}_j(0)]^2 \right> $.
Shown in Fig.~\ref{fig2}(c) is $\left< \delta r^2(t) \right>$ for the poorly annealed glass, for three aging times, $\tau_{\mathrm{age}} = 100,$ 100K, 1000K.
There is a significant upturn in $\left< \delta r^2(t) \right>$ for $\tau_{\mathrm{age}} = 100$, but no significant upturn for the longer aging times.
However, $\left< \delta r^2(t) \right>$ depends significantly on the aging time.

We calculated $C_{\lambda}(k,t)$ over the time frame shown in Fig.~\ref{fig2}(c) for the glasses with the three different aging times and determined
$\Gamma_{\lambda}$ from the envelope fits.  Shown in Figs.~\ref{fig2}(a) and (b) is sound attenuation at these three aging times.
With increasing $\tau_{age}$ we observe the small $k$ scaling
of $\Gamma_{\lambda}(k)$ changes from the $k^{1.5}$ for $\tau_{age}=100$, for which the mean square displacement
$\left< \delta r^2(t)\right>$ exhibits a significant upturn at late times,
to $k^{2}$ for the glasses with
$\tau_{age}=100$K and $\tau_{age}=1000$K, for which there is little to no upturn in  $\left< \delta r^2(t) \right>$.
This  suggests  the $k^{1.5}$ scaling observed at small wavevectors in our poorly annealed glass is due to aging, and for a glass
with no measurable aging, sound attenuation scales as $k^{2}$ for small wavevectors.

The $k^{1.5}$ scaling is consistent with the scaling reported by Mizuno~\cite{Mizuno2019CondesMatterFiniteT,Mizuno2019arxivFiniteT}, but they report
that their glass is not undergoing aging. Mizuno studied a monodisperse Lennard-Jones systems, which is known to be a very
poor glass former as it easily crystalizes. For this reason it may be closer to an elastic instability that is
predicted to give rise to the
$k^{1.5}$ scaling of sound attenuation \cite{MarruzzoMarginalStability2013,FerranteMarginalStability2013}. However,
we find that aging can also result in $k^{1.5}$ scaling at small wavevectors.

Next we examine the stability dependence of sound attenuation by
examining the parent temperature $T_p$ dependence of sound attenuation. Here we fix
$\tau_{\mathrm{age}} = 1000$ and $T = 0.062$ while we vary $T_p$ for glasses corresponding to poorly annealed,
$T_p = 0.2$, to our most stable glass, $T_p = 0.062$.  We note that $\tau_{\mathrm{age}}$ does not statistically
modify our results for $T_p \le 0.1$ since we cannot run the molecular dynamics simulations long enough to
significantly age the system.

\begin{figure}[!t]
\centering
\includegraphics[width=0.38\textwidth]{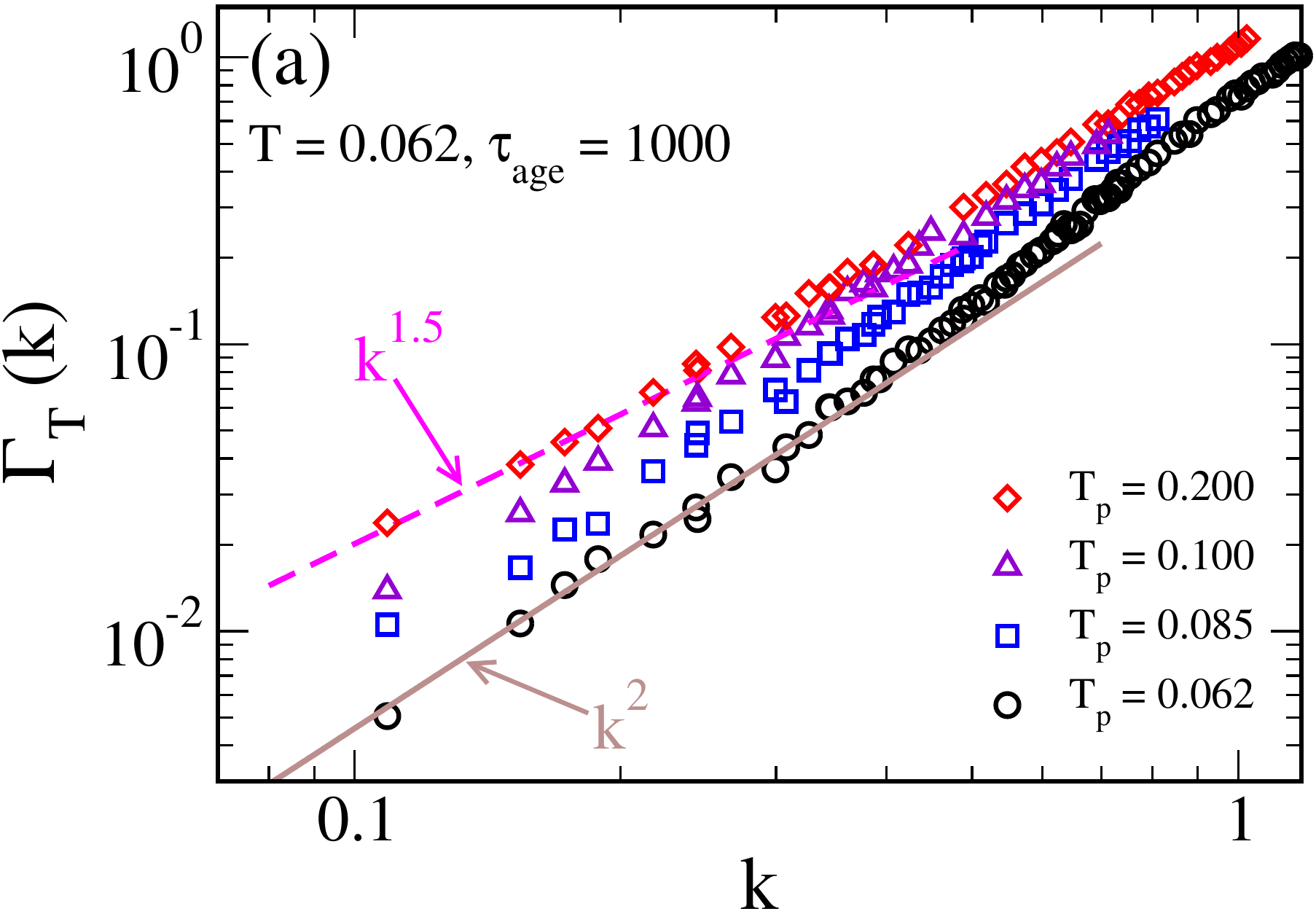}
\includegraphics[width=0.38\textwidth]{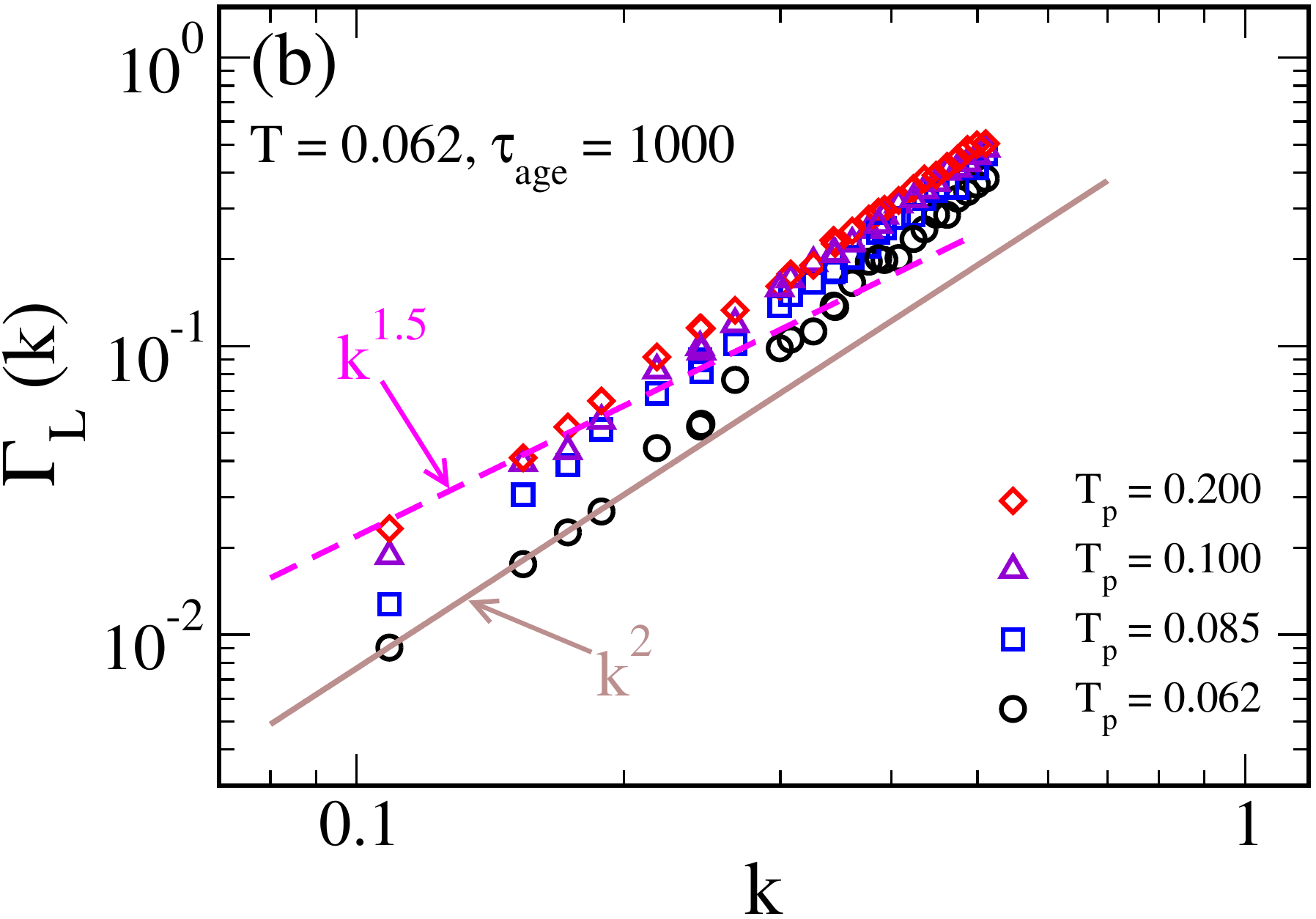}

\includegraphics[width=0.38\textwidth]{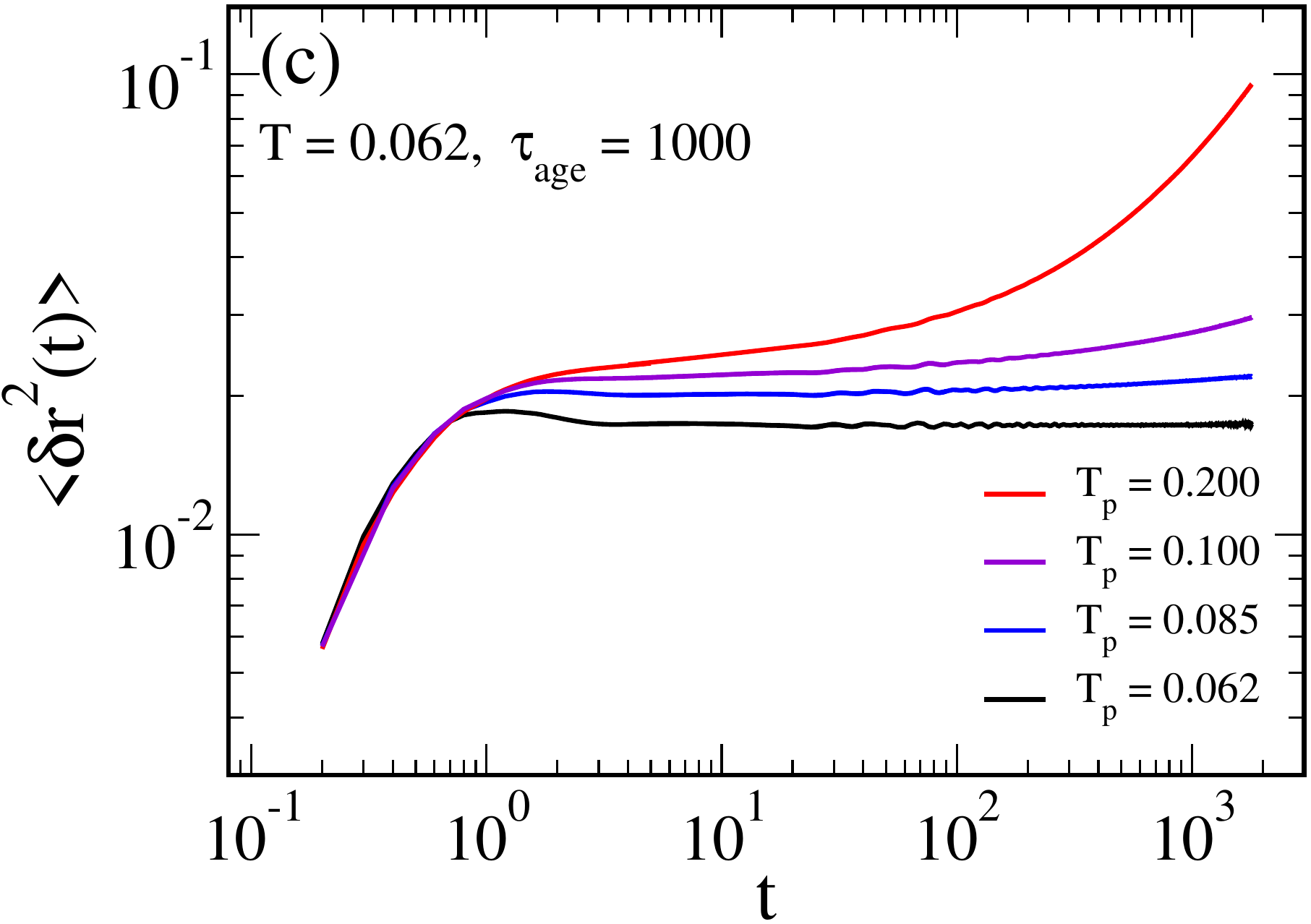}
\caption{  Wavevector $k$ dependence of  $\Gamma_T(k)$ (a) and $\Gamma_L(k)$ (b) in glasses with  the same temperature $T=0.062$ and aging time $\tau_{\mathrm{age}}=1000$ for different parent temperatures $T_p$.
The dashed and solid lines correspond to $k^{1.5}$ and $k^{2}$, respectively. (c) The mean-squared displacement $\langle \delta r^2(t)\rangle$ for glasses shown in (a) and (b).   }
\label{fig3}
\end{figure}

Shown in Fig.~\ref{fig3}(c) is $\left< \delta r^2(t) \right>$ for $T_p$ = 0.2, 0.1, 0.085, and 0.062.
There is a significant upturn for $T_p = 0.2$, but not for the other glasses. As shown previously,
this upturn corresponds to $\Gamma_{\lambda} \sim k^{1.5}$ at small wavevectors, Figs.~\ref{fig3}(a) and (b)
for the glass created at $T_p = 0.2$. We find that for the other glasses that $\Gamma_{\lambda} \sim k^2$
at small wavevectors.

Recent simulation studies~\cite{Wang2019SMattenuation,Lerner2019JCP} within
the harmonic approximation have concluded that sound attenuation decreases
with increasing glass stability. By studying sound attenuation at finite temperatures, we find the same conclusion also holds
when anharmonic effects are included. This is
supported by the observation that $\Gamma_{\lambda}(k)$ in our poorly annealed glass with $T_p=0.2$ is much larger than that in our exceptionally  stable  glass
with $T_p=0.062$. Recent studies~\cite{Schirmacher2007,Wang2019SMattenuation,Lerner2019JCP} have indicated that sound attenuation in the harmonic approximation may be proportional to the density of quasi-localized modes,
and that the density of these quasi-localized modes decreases rapidly with increasing stability~\cite{Wang2019NC,Rainone2019arxiv}. These modes may also be responsible for finite temperature
sound attenuation. It has also been found that the distribution of the local elastic constants narrows with increasing stability~\cite{Ali2020SM}, and this may
also give rise to the difference in the stability dependence of sound attenuation.

\section{Conclusions}

We examined finite temperature sound attenuation in glasses over a wide range of stabilities, with the most stable glasses having stability comparable to
that of experimental glasses.
For glasses undergoing aging, sound attenuation $\Gamma_{\lambda}(k)$ scales with wavevector $k$ as $k^{1.5}$.
We identify simulations where this aging effect is present by the appearance of an upturn in the mean square displacement.
When we see no upturn in the mean square displacement, then $\Gamma_{\lambda}(k) \sim k^2$ for small wavevectors.
As the glass's stability increases, through aging or by using a different preparation protocol,
sound attenuation decreases.  The decrease is most significant at small wavevectors.

For our most stable glasses at finite temperatures, we were able to clearly observe the three scaling regimes discussed by
Baldi \textit{et al.}\, \cite{Rayleigh_BaldiPRL2014}, the small and large wavevector quadratic scaling
and the intermediate wavevector quartic scaling. With increasing temperature, the small
wavevector sound attenuation increases significantly, while the large wavevector attenuation remains nearly unchanged. We find that the
wavevector (frequency) where thermal effects begin to become significant scales as $T^{0.22}$. It would be interesting to see if
this scaling is universal or depends on the specifics of the glass.

We determined that the coefficient
describing the small wavevector quadratic scaling increases as approximately $T^{1/2}$. This temperature dependence was predicted by
Schirmacher \cite{MarruzzoMarginalStability2013} and observed in simulations of Mizuno and Mossa \cite{Mizuno2019CondesMatterFiniteT,Mizuno2019arxivFiniteT}. However, according to Shirmacher's theory it should occur for a system close to an elastic instability where
the small wavevector scaling of sound attenuation is $k^{1.5}$ instead of the observed $k^2$.

Future work needs to examine the
role of quasi-localized modes and variations of local elasticity in sound attenuation. Mizuno determined
that the increase in attenuation with temperature correlated with an increase in the width of the distribution of local elastic constants
\cite{Mizuno2019arxivFiniteT}.
Shakerpoor {\it et al.}~\cite{Ali2020SM} found that the variation of the local elastic constants decreases with increasing stability for the glass
former examined in this work, and this decrease also correlates with the decrease in sound attenuation
reported by Wang \textit{et al}\cite{Wang2019SMattenuation}.

It has been argued by Schirmacher {\it et al.}~\cite{Schirmacher2007} on the basis of
 fluctuating elasticity theory that sound attenuation is related  to the excess density of states $D_{ex}(\omega)$,
 which corresponds to the density of quasi-localized modes in recent work~\cite{Mizuno2017PNAS,Lerner2016PRL,Wang2019NC}.
 Recent studies \cite{Buchenau1992,Wang2019NC,Lerner2016PRL,Mizuno2017PNAS,Kapteijns2018,Angelani2018,Schober1996,
 Gurevich2003,Benetti2018,HIkeda2019,Stanifer2018,Rainone2019arxiv}
 within the harmonic approximation show that $D_{ex}(\omega)$ scales universally with $\omega^4$ at low $\omega$
 and that $D_{ex}(\omega) \sim \Gamma(\omega)$ \cite{Wang2019SMattenuation}.
 Future work should examine the connection between the change of sound attenuation and $D_{ex}(\omega)$, specifically
 whether the anharmonicity also alters the $\omega^4$ scaling of $D_{ex}(\omega)$ and the connection to
 the anharmonic properties of quasi-localized modes~\cite{Mizuno2019Anharm, Xu2010EPL}.

\section*{Conflicts of interest}
There are no conflicts to declare.

\section*{Acknowledgements}
We thank H. Mizuno and his coworkers for kind correspondence regarding some results of this work.
This work was supported by  NSF
Grants DMR-1608086 (E.F. and G.S) and CHE-1800282 (E.F. and G.S.), and the Start-up Fund from Anhui University 
S020318001/02 (L.W.). We also acknowledge Beijing Computational Science Research Center and the High-Performance Computing Platform of Anhui University for providing computing resources.

\end{document}